\documentclass[12pt,thmsa]{article}

\begin{document}

\title{Fractional spin through quantum affine algebras with vanishing central
charge.}
\author{M. Mansour \thanks{%
mansour70@mailcity.com} and E. H. Zakkari \thanks{%
hzakkari@yahoo.fr} \\
Laboratory of Theoretical Physics\\
University Mohamed V\\
PO BOX 1014\\
Rabat, Morocco.}
\date{}
\maketitle

\begin{abstract}
In this paper, we study the fractional decomposition of the quantum
enveloping affine algebras $U_Q(\hat A(n))$ and $U_Q(\widehat{C}(n))$ with
vanishing central charge in the limit $Q\rightarrow q=e^{\frac{2i\pi }k}$ .
This decomposition is based on the bosonic representation and can be related
to the fractional supersymmetry and $k$-fermionic spin. The equivalence
between the quantum affine algebras and the classical ones in the fermionic
realization is also established.
\end{abstract}

\newpage\

\section{Introduction}

The concept of quantum group and algebras $[1,2]$, have enriched the arena
of mathematics and theoretical physics. Quantum groups appeared in studying
Yang-Baxter equations $[3]$ as well as scattering method $[4]$. In $[5,6]$
the quantum analogue of Lie superalgebras was constructed. The quantized
enveloping algebras associated to affine algebras and superalgebras are
given in $[1,7]$. It is well known that the boson realization is a very
powerful and elegant method for studying quantum algebras representations.
Based on this method, the representation theory of quantum affine algebras
has been an object of intensive studies, namely, the results for the
oscillator representations of affine algebras. There are obtained $[8-10]$
through consistent realization involving deformed Bose and Fermi operators $%
[11,12]$.

To make a connection with the quantum group theory, a new geometric
interpretation of fractional supersymmetry has been introduced in $[13-17]$.
In these papers, the authors show that the one-dimensional superspace is
isomorphic to the braided line when the deformation parameter goes to a root
of unity. The similar technics are used, in $[18]$, to show how internal
spin arises naturally in a certain limit of the $Q$-deformed momentum
algebras $U_Q(sl(2))$.

Indeed, using $Q$-Schwinger realization, it is proved that the decomposition
of the $U_Q(sl(2))$ into a direct product $U(sl(2))$ and the deformed $%
U_q(sl(2))$ $($ note that $U_Q(sl(2))=$ $U_q(sl(2))$ at $Q=q$ $)$. The
property of splitting quantum algebras $A_n,$ $B_n,$ $C_n$ and $D_n$ and
quantum superalgebras $C(n),$ $B(n,m),$ $C(n+1)$ and $D(n,m)$ in the limit $%
Q\rightarrow q$ is investigated in $[19]$.

We also notice that the case of deformed Virasoro algebras and some other
particular quantum (Super) algebras is given in $[20]$.

The aim of this paper is to investigate the decomposition property of the
quantum affine algebras with vanishing central charge $U_Q(\hat A(n))$ and $%
U_Q(\hat C(n))$ in the limit $Q\rightarrow q$ . We start in section $2$ by
defining $k$-fermionic algebra. In section $3$, we discuss the decomposition
property of $Q$-boson oscillator in the limit $Q\rightarrow q$ . We
introduce the way in which one obtains two independent objects, an ordinary
boson and a $k$-fermion, from a $Q-$deformed boson when $Q$ goes to the root
of unity $q$. We also establish the equivalence between a $Q$-deformed
fermion and conventional (ordinary) one. Using these results, we analyze the
limit $Q\rightarrow q$ of the quantum affine algebras with vanishing central
charge $U_Q(\hat A(n))$ (section 4) and $U_Q(\widehat{C}(n))$ (section 5).
Some concluding remarks are given in section $6.$

\section{Preliminaries about k-fermionic algebra.}

The $q$-deformed bosonic algebra $\Sigma _q$ generated by $A^{+}$, $A^{-}$
and number operator $N$ is given by:

\begin{equation}
A^{-}A^{+}-qA^{+}A^{-}=q^{-N}
\end{equation}

\begin{equation}
A^{-}A^{+}-q^{-1}A^{+}A^{-}=q^N
\end{equation}

\begin{equation}
q^NA^{\pm }q^{-N}=q^{\pm 1}A^{\pm }
\end{equation}

\begin{equation}
q^Nq^{-N}=q^{-N}q^N=1,
\end{equation}
where the deformation parameter:

\begin{equation}
q=e^{\frac{2i\pi }l},\; l\in N-\{0,1\},
\end{equation}

is a root of unity.

The annihilation operator $A^{-}$ is hermitian conjugate to creation
operator $A^{+}$ and $N$ is hermitian also. From equations $(1)-(4)$, it is
easy to have the following relations:

\begin{equation}
A^{-}(A^{+})^n=[[n]]q^{-N}(A^{+})^{n-1}+q^n(A^{+})^nA^{-}
\end{equation}

\begin{equation}
(A^{-})^nA^{+}=[[n]](A^{-})^{n-1}q^{-N}+q^nA^{+}(A^{-})^n,
\end{equation}
where the notation $[[$ $]]$ is defined by:

\begin{equation}
\lbrack [n]]=\frac{1-q^{2n}}{1-q^2}
\end{equation}

We introduce a new variable $k$ defined by:
\begin{equation}
k=l\;\; \hbox{for  odd  values  of}\;\; l,
\end{equation}

\begin{equation}
k\,=\,\frac l2\;\;\hbox{for even values of}\;\;l,
\end{equation}
such that for odd $l$ (resp. even $l$ ), we have $q^k=1$ $($resp. $q^k=-1).$
In the particular case $n=k$, equations $(6)-(7)$ permit us to have:

\begin{equation}
A^{-}(A^{+})^k=\pm (A^{+})^kA^{-}
\end{equation}

\begin{equation}
(A^{-})^kA^{+}=\pm A^{+}(A^{-})^k,
\end{equation}
and the equations $(1)-(5)$ yield to:

\begin{equation}
q^N(A^{+})^k=(A^{+})^kq^N
\end{equation}

\begin{equation}
q^N(A^{-})^k=(A^{-})^kq^N
\end{equation}
One can show that the elements $(A^{-})^k$ and $(A^{+})^k$ are the elements
of the centre of $\sum_q$ algebra (odd values for $l$); and the irreducible
representations are $k$-dimensional. These two properties lead to:
\begin{equation}
(A^{+})^k=\alpha I
\end{equation}

\begin{equation}
(A^{-})^k=\beta I.
\end{equation}

The extra possibilities parameterized by:

\[
(1) \;\; \alpha \, =\, 0, \;\; \beta \neq 0
\]

\[
(2) \;\; \alpha \neq 0, \;\; \beta =0
\]

\[
(3)\;\;\alpha \neq 0,\;\;\beta \neq 0,
\]
are not relevant for the considerations of this paper. In the two cases $(1)$
and $(2)$ we have the so-called semi-periodic (semi-cyclic) representation
and the case $(3)$ correspond to the periodic one. In what follows, we are
interested in a representation of the algebra $\sum_q$ such that the
following:

\[
(A^{\mp })^k=0,
\]
is satisfied. We note that the algebra $\sum_{-1}$ obtained for $k=2$,
correspond to ordinary fermion operators with $(A^{+})^2=0$ and $(A^{-})^2=0$
which reflects the exclusion's Pauli principle. In the limit case where $%
k\rightarrow \infty $, the algebra $\sum_1$ correspond to the ordinary
bosons. For other values of $k$, the $k$-fermions operators interpolate
between fermions and bosons, these are also called anyons with fractional
spin in the sense of Majid $[21,22]$.

\section{Fractional spin through Q-boson.}

In the previous section, we have worked with $q$ at root of unity. In this
case, quantum oscillator $(k$-fermionic$)$ algebra exhibit a rich
representation with very special properties different from the case where $q$
is generic. So, in the first case the Hilbert space is finite dimensional.
In contrast, where $q$ is generic, the Fock space is infinite dimensional.
In order to investigate the decomposition of $Q$-deformed boson in the limit
$Q\rightarrow e\frac{2i\pi }k$ we start by recalling the $Q$-deformed
algebra $\Delta _Q$.

The algebra $\Delta _Q$ generated by an annihilation operator $B^{-}$, a
creation operator $B^{+}$ and a number operator $N_B$:

\begin{equation}
B^{-}B^{+}-QB^{+}B^{-}=Q^{-N_B}
\end{equation}

\begin{equation}
B^{-}B^{+}-Q^{-1}B^{+}B^{-}=Q^{N_B}
\end{equation}

\begin{equation}
Q^{N_B}B^{+}Q^{-N_B}=QB^{+}
\end{equation}

\begin{equation}
Q^{N_B}B^{-}Q^{-N_B}=Q^{-1}B^{-}
\end{equation}

\begin{equation}
Q^{N_B}Q^{-N_B}=Q^{-N_B}Q^{^{+}N_B}=1.
\end{equation}

From the above equations, we obtain:
\begin{equation}
\lbrack
Q^{-N_B}B^{-},[Q^{-N_B}B^{-},[....[Q^{-N_B}B^{-},(B^{+})^k]_{Q^{2k}}...]_{Q^4}]_{Q^2}]=Q^{%
\frac{k(k-1)}2}[k]!
\end{equation}
where the $Q$-deformed factorial is given by:

\begin{equation}
\lbrack k]!=[k][k-1][k-2]...............[1],
\end{equation}
and:

\[
\lbrack 0]!=1
\]

\[
\lbrack k]=\frac{Q^k-Q^{-k}}{Q-Q^{-1}}\hbox{ .}
\]

The $Q$-commutator, in equation $(22)$, of two operators $A$ and $B$ is
defined by:

\[
\lbrack A,B]_Q=AB-QBA
\]

The aim of this section is to determine the limit of $\Delta _Q$ algebra
when $Q$ goes to the root of unity $q$. The starting point is the limit $%
Q\rightarrow q$ of the equation $(22),$

\[
\lim_{Q\rightarrow q}\frac
1kQ^{-N_B}[Q^{-N_B}B^{-},[Q^{-N_B}B^{-},[....[Q^{-N_B}B^{-},(B^{+})^k]_{Q^{2k}}...]_{Q^4}]_{Q^2}]
\]

\begin{equation}
=\lim_{Q\rightarrow q}\frac{Q^{\frac{k(k-1)}2}}{[k]!}[%
Q^{-N_B}(B^{-})^k,(B^{+})^k]=q^{\frac{k(k-1)}2}
\end{equation}

This equation can be reduced to:

\begin{equation}
\lim_{Q\rightarrow q}[\frac{Q^{\frac{kN_B}2}(B^{-})^k}{([k]!)^{\frac 12}},%
\frac{(B^{+})^kQ^{\frac{kN_B}2}}{([k]!)^{\frac 12}}]=1.
\end{equation}

Since $q$ is a root of unity, it is possible to change the sign on the
exponent of $q^{\frac{kN_B}2}$ terms in the above equation.

We define the operators as in $[18]$:

\begin{equation}
b^{-}=\lim_{Q\rightarrow q}\frac{Q^{\pm \frac{kN_B}2}}{([k]!)^{\frac 12}}%
(B^{-})^k,\,b^{+}=\lim_{Q\rightarrow q}\frac{(B^{+})^kQ^{^{\pm }\frac{kN_B}2}%
}{([k]!)^{\frac 12}},
\end{equation}
which lead to an ordinary boson algebra noted $\Delta _0$, generated by:

\begin{equation}
\lbrack b^{-},b^{+}]=1.
\end{equation}

The number operator of this new bosonic algebra defined as the usual case, $%
N_b=b^{+}b^{-}$. At this stage we are in a position to discuss the splitting
of $Q$-deformed boson in the limit $Q\rightarrow q$ . Let us introduce the
new set of generators given by:

\begin{equation}
A^{-}=B^{-}q^{-\frac{kN_b}2}
\end{equation}

\begin{equation}
A^{+}=B^{+}q^{-\frac{kN_b}2}
\end{equation}

\begin{equation}
N_A=N_B-kN_b,
\end{equation}
which define a $k$-fermionic algebra:

\begin{equation}
\lbrack A^{+},A^{-}]_{q^{-1}}=q^{N_A}
\end{equation}

\begin{equation}
\lbrack A^{-},A^{+}]_q=q^{-N_A}
\end{equation}

\begin{equation}
\lbrack N_A,A^{\pm }]=\pm A^{\pm }.
\end{equation}
It is easy to verify that the two algebras generated by the set of operators
$\{b^{+},b^{-},N_b\}$ and $\{A^{+},A^{-},N_A\}$ are mutually commutative. We
conclude that in the limit $Q\rightarrow q$ , the $Q$-deformed bosonic
algebra oscillator decomposes into two independent oscillators, an ordinary
boson and $k$-fermion; formally one can write:

\[
\lim_{Q\rightarrow q}\Delta _Q\equiv \Delta _0\otimes \Sigma _q,
\]
where $\Delta _0$ is the classical bosonic algebra generated by the
operators $\{b^{+},b^{-},$ $N_b\}.$

Similarly, we want to study the $Q$-fermion algebra at root of unity. To do
this, we start by considering the $Q$- deformed fermionic algebra, noted $%
\Xi _Q$:

\begin{equation}
F^{-}F^{+}+QF^{+}F^{-}=Q^{N_{F}}
\end{equation}

\begin{equation}
F^{-}F^{+}+Q^{-1}F^{+}F^{-}=Q^{-N_F}
\end{equation}

\begin{equation}
Q^{N_F}F^{+}Q^{-N_F}=QF^{+}
\end{equation}

\begin{equation}
Q^{N_F}F^{-}Q^{-N_F}=Q^{-1}F^{-}
\end{equation}

\begin{equation}
Q^{N_F}Q^{-N_F}=Q^{-N_F}Q^{N_F}=1
\end{equation}

\begin{equation}
(F^{+})^2=0, \; \; (F^{-})^2=0
\end{equation}

We define the new fermionic operators as follow:

\begin{equation}
f^{+}=\lim_{Q\rightarrow q}F^{+}Q^{^{-}\frac{N_F}2}
\end{equation}

\begin{equation}
f^{-}=\lim_{Q\rightarrow q}Q^{^{-}\frac{N_F}2}F^{-}.
\end{equation}

By a direct calculus, we obtain the following anti-commutation relation:

\begin{equation}
\{f^{-},\, f^{+}\}=1.
\end{equation}

Moreover, we have the nilpotency condition:

\begin{equation}
(f^{-})^2=0, \;\;(f^{+})^2=0.
\end{equation}

Thus, we see that the $Q$-deformed fermion reproduce the conventional
(ordinary) fermion. The same convention notation permits us to write:

\[
\lim_{Q\rightarrow q}\Xi _Q\equiv \Sigma _{-1}
\]

\section{Quantum affine algebra $U_Q(\hat A(n))$ at $Q$ a root of unity}

We apply the above results to derive the property of decomposition of
quantum affine algebra with vanishing central charge $U_Q(\hat A(n))$ in the
limit$Q\rightarrow q$ . Recalling that the $U_Q(\hat A(n))$ algebra is
generated by the set of generators $\{e_i,$ $f_i,$ $h_i,$ $0\leq i$ $\langle
$ $n\}$ satisfying the following relations:

\begin{equation}
\lbrack e_i,f_j]=\delta _{ij}\frac{Q_i^{h_i}-Q_i^{-h_i}}{Q_i-Q_i^{-1}}
\end{equation}

\begin{equation}
\lbrack h_i,e_j]=a_{ij}e_j;\hspace{1.0in}[f_i,h_j]=a_{ij}f_j
\end{equation}

\begin{equation}
\lbrack h_i,h_j]=[e_i,e_j]=[f_i,f_j]=0.
\end{equation}

The quantum affine algebra with vanishing central charge $U_Q(\hat A_n)$
admits two $Q$-oscillators representations: bosonic and fermionic ones; in
the bosonic realization, the generators of $U_Q(\hat A_n)$ can be
constructed by introducing $(n+1)$ $Q$-deformed bosons as follows:

\[
e_{i}=B_{i}^{-}B_{i+1}^{+},1\leq i\leq n
\]

\[
f_{i}=B_{i}^{+}B_{i+1}^{-},1\leq i\leq n
\]

\[
h_i=-N_i+N_{i+1},1\leq i\leq n
\]

\[
e_{0}=B_{n+1}^{-}B_{1}^{+}
\]

\[
f_{0}=B_{1}^{-}B_{n+1}^{+}
\]

\[
h_0=N_1-N_{n+1}.
\]

The fermionic realization of $U_Q(\hat A(n))$ is given by:

\[
e_i=F_i^{+}F_{i+1}^{-},1\leq i\leq n
\]

\[
f_i=F_i^{-}F_{i+1}^{+},1\leq i\leq n
\]

\[
h_i=N_i-N_{i+1},1\leq i\leq n
\]

\[
e_0=F_{n+1}^{+}F_1^{-}
\]

\[
f_0=F_1^{+}F_{n+1}^{-}
\]

\[
h_0=-N_1+N_{n+1}.
\]

At this stage, our aim is to investigate the limit $Q\rightarrow q$ of the
affine algebra with vanishing central charge $U_Q(\hat A_n)$. As it is
already mentioned in the introduction, our analysis is based on the $Q$%
-oscillator representation based on $Q$-Schwinger realization. In the limit $%
Q\rightarrow q$, the splitting of $Q$-deformed bosons leads to classical
bosons $\{b_i^{+},b_i^{-},N_{b_i},$ $1\leq i\leq n\}$ given by the equations$%
(26)-$ $(27)$ and $k$-fermionic algebra $\{A_i^{+},A_i^{-},N_{A_i},$ $1\leq
i\leq n\}$given by equations$(31)-(33)$. From the classical bosons, we
define for $i=1,...,n$ the operators:

\begin{equation}
e_i=b_i^{-}b_{i+1}^{+}
\end{equation}

\begin{equation}
f_i=b_i^{+}b_{i+1}^{-}
\end{equation}

\begin{equation}
h_i=-N_{b_i}+N_{b_{i+1}}
\end{equation}

\begin{equation}
e_0=b_1^{-}b_{n+1}^{+}
\end{equation}

\begin{equation}
f_0=b_1^{+}b_{n+1}^{-}
\end{equation}

\begin{equation}
h_0=-N_{b_1}+N_{b_{n+1}},
\end{equation}
the set $\{e_i,f_i,h_i,0\leq i\leq n\}$ generate the classical algebra $%
U(\hat A(n)).$ From the remaining generators $\{A_i^{+},A_i^{-},N_{A_i},$ $%
1\leq i\leq n+1\}$, we can realize $U_q(\hat A(n))$, generated by $E_i,$ $%
F_i,$ $H_i,E_0,F_0$ and $H_0$ where:

\begin{equation}
E_i=A_i^{-}A_{i+1}^{+},1\leq i\leq n
\end{equation}

\begin{equation}
F_i=A_i^{+}A_{i+1}^{-},1\leq i\leq n
\end{equation}

\begin{equation}
H_i=-N_{A_i}+N_{A_{n+1}},1\leq i\leq n
\end{equation}

\begin{equation}
E_{0}=A_{1}^{+}A_{n+1}^{-}
\end{equation}

\begin{equation}
F_{0}=A_{1}^{-}A_{n+1}^{+}
\end{equation}

\begin{equation}
H_0=N_{A_1}-N_{A_{n+1}}.
\end{equation}
The algebra $U_q(\hat A(n))$ is the same version of $U_Q(\hat A_n)$ obtained
by simply taking $Q=q$ and $B_i\sim A_i.$ Due to the commutativity of
elements of $U_q(\hat A(n))$ and $U(\hat A_n),$ we obtain the following
decomposition of the quantum affine algebra $U_Q(\hat A_n)$ in the bosonic
realization

\[
\lim_{Q\rightarrow q}U_Q(\hat A_n)\equiv U_q(\hat A(n))\otimes U(\hat A(n)).
\]
We discuss now the equivalence between $U_Q(\hat A_n)$ and $U(\hat A(n))$
algebras in the fermionic realization. Indeed, we have discussed in section $%
2$, how one can identify the conventional fermions with $Q$-deformed
fermions. Consequently, due to this equivalence, it is possible to construct
$Q$-deformed affine algebras $U_Q(\hat A_n)$ using ordinary fermions. It is
also possible to construct the affine algebra $U(\hat A_n)$ by considering $%
Q $-deformed fermions. So, in the fermionic realization we have equivalence
between $U(\hat A_n)$ and $U_Q(\hat A_n).$ To be more clear, we consider the
$U_Q(\hat A_n)$ in the $Q$-fermionic representation. Where the generators
are given by:

\begin{equation}
e_i=F_i^{-}F_{i+1}^{+},1\leq i\leq n
\end{equation}

\begin{equation}
f_i=F_i^{+}F_{i+1}^{-},1\leq i\leq n
\end{equation}

\begin{equation}
h_i=N_{F_i}-N_{F_{i+1}},1\leq i\leq n
\end{equation}

\begin{equation}
e_0=F_{n+1}^{+}F_1^{-}
\end{equation}

\begin{equation}
f_{0}=F_{1}^{+}F_{n+1}^{-}
\end{equation}

\begin{equation}
h_0=-N_{F_1}+N_{F_{n+1}}.
\end{equation}

Due to the equivalence fermion $Q$-fermion, the operators $f_i^{-}$, $%
f_i^{+} $ are defined as a constant multiple of conventional fermion
operators:

\begin{equation}
f_i^{+}=F_i^{+}Q^{\frac{-N_{Fi}}2}
\end{equation}

\begin{equation}
f_i^{-}=Q^{\frac{-N_{F_i}}2}F_i^{-},
\end{equation}
from which we can realize the generators:

\begin{equation}
E_i=f_i^{-}f_{i+1}^{+},\, 1\leq i\leq n
\end{equation}

\begin{equation}
F_i=f_i^{+}f_{i+1}^{-},\, 1\leq i\leq n
\end{equation}

\begin{equation}
H_i=N_{f_i}-N_{f_{i+1}},\, 1\leq i\leq n
\end{equation}

\begin{equation}
E_0=f_{n+1}^{+}f_1^{-}
\end{equation}

\begin{equation}
F_{0}=f_{1}^{+}f_{n+1}^{-}
\end{equation}

\begin{equation}
H_0=-N_{f_1}+N_{f_{n+1}}.
\end{equation}

The set $\{E_i,F_i,H_i$ $;$ $0\leq i\leq n\}$ generate the classical affine
algebra $U(\hat A_n)$ in the fermionic representation and we have

\[
U_q(\hat A(n))\equiv U(\hat A(n)).
\]

\section{Quantum affine algebra $U_Q(\widehat{C}(n))$ at a root of unity.}

Let $Q\in C-\{0\}$ be the deformation parameter. The quantum affine algebra $%
U_Q(\widehat{C}(n))$ is described in the Serre-Chevalley basis in terms of
the simple root $e_i$, $f_i$ and Cartan generators $h_i$, where $i=0,...n$,
satisfy the following commutation relations:

\begin{equation}
\lbrack e_i,f_j]=\delta _{ij}\frac{Q^{h_i}-Q^{-h_i}}{Q_i-Q_i^{-1}}
\end{equation}

\begin{equation}
\lbrack e_i,e_j]=[f_i,f_j]=[h_i,h_j]=0
\end{equation}
\begin{equation}
\lbrack h_i,e_j]=a_{ij}e_j,\,[h_i,f_j]=-a_{ij}f_j.
\end{equation}

An explicit realization of the quantum affine symplectic algebra $U_Q(%
\widehat{C}_{\,}(n))$ has been given by L.Frappat et al in $[23].$ In the
particular case of the quantum affine symplectic algebra with vanishing
central charge $U_Q(\widehat{C}_{\,}(n))$; the generators can be realized in
the bosonic case by:

\begin{equation}
e_{i}=B_{i}^{+}B_{i+1}^{-}+B_{2n-i}^{+}B_{2n-i+1}^{-},1\leq i\leq n-1
\end{equation}

\begin{equation}
f_{i}=B_{i}^{-}B_{i+1}^{+}+B_{2n-i}^{-}B_{2n-i+1}^{+},1\leq i\leq n-1
\end{equation}

\begin{equation}
h_{i}=N_{B_{i}}-N_{B_{i+1}}+N_{B_{2n-i}}-N_{B_{2n-i+1}},1\leq i\leq n-1
\end{equation}

\begin{equation}
e_{n}=B_{n+1}^{-}B_{n}^{+}
\end{equation}

\begin{equation}
f_{n}=B_{n+1}^{+}B_{n}^{-}
\end{equation}

\begin{equation}
h_{n}=N_{B_{n}}-N_{B_{n+1}}
\end{equation}

\begin{equation}
e_{0}=B_{2n}^{+}B_{1}^{-}
\end{equation}

\begin{equation}
f_{0}=B_{1}^{+}B_{2n}^{-}
\end{equation}

\begin{equation}
h_0=N_{B_{2n}}-N_{B_1}.
\end{equation}

Due to the property of $Q$-boson decomposition in the $Q\rightarrow q$
limit, each $Q$-boson $\{B_i^{-},B_i^{+},N_{B_i}\}$ reproduce an ordinary
bosonic algebra $\{b_i^{-},b_i^{+},N_{b_i}\}$ and $k$-fermion operators $%
\left\{ A_i^{-},A_i^{+},N_{A_i}\right\} .$

From the set $\left\{ b_i^{+},\, b_i^{-},\, N_{b_i},\, i=0....n\right\} $ we
can construct the classical affine algebra $U(\hat C(n))$ as follow:

\begin{equation}
E_i=b_i^{+}b_{i+1}^{-}+b_{2n-i}^{+}b_{2n-i+1}^{-},\, 1\leq i\leq n-1
\end{equation}

\begin{equation}
F_i=b_i^{-}b_{i+1}^{+}+b_{2n-i}^{-}b_{2n-i+1}^{+},\, 1\leq i\leq n-1
\end{equation}

\begin{equation}
H_i=N_{b_i}-N_{b_{i+1}}+N_{b_{2n-i}}-N_{b_{2n-i+1}},\, 1\leq i\leq n-1
\end{equation}

\begin{equation}
E_{n}=b_{n+1}^{-}b_{n}^{+}
\end{equation}

\begin{equation}
F_{n}=b_{n+1}^{+}b_{n}^{-}
\end{equation}

\begin{equation}
H_{n}=N_{b_{n}}-N_{b_{n+1}}
\end{equation}

\begin{equation}
E_{0}=b_{2n}^{+}b_{1}^{-}
\end{equation}

\begin{equation}
F_{0}=b_{1}^{+}b_{2n}^{-}
\end{equation}

\begin{equation}
H_0=N_{b_{2n}}-N_{b_1}.
\end{equation}

From the $k$-fermionic operators $\left\{ A_i^{-},\,A_i^{+},\,N_{A_i},1\leq
i\leq n+1\right\} $, one can construct as in equations $(76)-(84)$ the $q$%
-deformed affine algebra $U_q(\hat C(n))$. It is easy to verify that $U_q(%
\widehat{C}(n))$ and $U(\widehat{C}(n))$ are mutually commutative. As a
result, we have the following decomposition of quantum algebra $U_Q(\widehat{%
C}(n))$ in the limit $Q\rightarrow q$ :

\[
\lim_{Q\rightarrow q}U_Q(\widehat{C}(n))\equiv U(\widehat{C}(n))\otimes U_q(%
\widehat{C}(n)).
\]
The equivalence between $U_Q(\widehat{C}(n))$ and $U(\widehat{C}(n))$
algebras in the fermionic representation can be easily deduced; in fact we
can construct the affine deformed algebra $U_Q(\widehat{C}(n))$ using the
ordinary fermions and conversely, the classical affine algebra $U(\widehat{C}%
(n))$ can be realized in terms of deformed fermions. Indeed, we consider the
$U_Q(\widehat{C}(n))$ in the $Q$-fermionic representation, where the
generators are given by:

\begin{equation}
E_i=F_i^{+}F_{i+1}^{-}+F_{2n-i}^{+}F_{2n-i+1}^{-},\, 1\leq i\leq n-1
\end{equation}

\begin{equation}
F_i=F_i^{-}F_{i+1}^{+}+F_{2n-i}^{-}F_{2n-i+1}^{+},\, 1\leq i\leq n-1
\end{equation}

\begin{equation}
H_i=N_{B_{i+1}}-N_{B_i}+N_{B_{_{2n-i+1}}}-N_{B_{2n-i}}^{},\,1\leq i\leq n-1
\end{equation}

\begin{equation}
E_{n}=F_{n+1}^{-}F_{n}^{+}
\end{equation}

\begin{equation}
F_{n}=F_{n+1}^{+}F_{n}^{-}
\end{equation}

\begin{equation}
H_n=N_{B_{n+1}}-N_{B_n}^{}
\end{equation}

\begin{equation}
E_{0}=F_{2n}^{+}F_{1}^{-}
\end{equation}

\begin{equation}
F_{0}=F_{1}^{+}F_{2n}^{-}
\end{equation}

\begin{equation}
H_0=N_1-N_{B_{2n}}^{}.
\end{equation}

As in the case of $U_Q(\hat A_n)$, the $Q-$deformed fermions \ can be
identified to classical ones.

So, we can deduced that in the fermionic representation the $Q-$deformed
algebra $U_Q(\widehat{C}(n))$ is equivalent to the classical affine algebra $%
U(\widehat{C}(n))$ and one can write:

\[
\lim_{Q\rightarrow q}U_Q(\widehat{C}(n))\equiv U(\widehat{C}(n)).
\]

\section{Conclusion}

In this paper we have worked with $q$ at root of unity. In this case,
quantum oscillator $(k$-fermionic$)$ algebra exhibit a rich representation
with very special properties different from the case where $q$ is generic.We
have presented the general method leading to the investigation of the limit $%
Q\rightarrow q=e^{\frac{2i\pi }k}$ of the quantum affine algebras with
vanishing central charge $U_Q(\hat A_n)$ and $U_Q(\widehat{C}(n))$. We note
that the $Q$-oscillator representation is crucial in this manner of
splitting in this paper. The technics and formulae used in this paper, will
be useful to extend this study to the infinite deformed algebras $[24]$, and
quantum affine superalgebras $[25].$

\newpage\

\section*{References}

$[1]$ V.G Drinfeld, Proc. Int. Cong. Math. (Berkley,1986), Vol 1. p. 798.

\noindent$[2]$ M. Jimbo, Lett. Math. Phys. 11 (1986) 247.

\noindent$[3]$ P. Kulish and E. Sklyanin, Lecture Notes in Physics, VoL 151 %
\\ (Springer,1981), p. 61.

\noindent$[4]$ L. D. Fadeev, les Houches Lectures, (Elsivier, 1982), p. 563.

\noindent$[5]$ R.Floreanini, P. Spridinov and L. Vinet, Phys. Lett. B (1990)
242 .

\noindent$[6]$ R. Floreanini, P. Spiridinov and L.Vinet, Commun. Math. 137
(1991) 149.

\noindent$[7]$ H. Yamane,\textit{\ }\thinspace Proceedings of the XX th
IGGTMP, Toyonaka (1994) ; World Scientific Singapore (1995).\

\noindent$[8]$ L. Frappat, A. Sciarrino, S. Sciuto and Sorba, Phys. Lett. B
(1996) 369 .

\noindent$[9]$ L. Frappat, A. Sciarrino, S. Sciuto and Sorba$,$ J.Phys A,
30, (1997) 903.

\noindent$[10]$ A. J. Feingold, I. B. frenkel. Ad. Math. 56 (1985) 117.

\noindent$[11]$ L. C. Biedenharn, J. Phys. A 22 (1998) L 873.

\noindent$[12]$ A. J. Macfarlane, J. Phys. A 22 (1988) 4581.

\noindent$[13]$ R. S. Dunne, A. J. Macfarlane, J. A. de Azcarraga, and J.C.
Perez Bueno, Phys. Lett B 387 (1996) 294.

\noindent$[14]$ R. S. Dunne, A. J. Macfarlane, J. A. de Azcarraga, and J.C.
Perez Bueno, hep-th/960087.

\noindent$[15]$ R. S. Dunne, A. J. Macfarlane, J. A. de Azcarraga, and J.C.
Perez Bueno, Czech. J. P. \ Phys. 46, (1996) 1145.

\noindent$[16]$ J. A. Azcarraga, R. S. Dunne, A. J. Macfarlane and
J.C. Perez Bueno, Czech. J. P. \ Phys. 46, (1996) 1235.

\noindent$[17]$ R. S. Dunne, J. Math. Phys 40, (1999) 1180 .

\noindent$[18]$ R. S. Dunne, hep-th/9703137.

\noindent$[19]$ M. Mansour, M. Daoud and Y. Hassouni, Phys. Lett. B (1999)
454 .

\noindent$[20]$ M. Mansour, M. Daoud and Y. Hassouni, Rep. Math. Phys. Vol.
44 (1999), 435.

\noindent$[21]$ S. Majid, proc. Of 2nd Max Born Symposium, Wroclam, Poland,
(1992), z.Oziewicz et al. Khluwwer.\ \ \ \ \

\noindent$[22]$ S. Majid, hep-th/9410241.

\noindent$[23]$ L.Frappat, A.Scarrino, S.Sciuto and P.Sorba; Czech.J.Phys
47, (1997) 5 .

\noindent$[24]$ M. Mansour, E. H. Zakkari, \textit{Q-Fractional spin through
some infinite deformed} \textit{algebra}, to be submitted.\ \ \ \ \

\noindent$[25]$ M. Mansour, work in progress.

\end{document}